\documentclass[prb,reprint,amsmath,amssymb,floatfix,citeautoscript,superscriptaddress]{revtex4-1}
\usepackage{cancel}
\usepackage{amsmath}
\usepackage{physics}
\usepackage{graphicx}
\usepackage{ulem}
\usepackage{amssymb}
\usepackage{sidecap}
\usepackage{amsthm}
\usepackage{bm}
\usepackage{dcolumn}% Align table columns on decimal point
\usepackage{braket}% Enables braket notation
\usepackage{longtable}
\usepackage{txfonts}

\usepackage{color}
\usepackage[usenames,dvipsnames]{xcolor}
\definecolor{myblue}{rgb}{0,0,1}
\usepackage[breaklinks=true,colorlinks=true,linkcolor=myblue,urlcolor=myblue,citecol  or=myblue]{hyperref}

\newcommand{\eps}{{\varepsilon}}
\newcommand{\gh}{{g_{\mathrm{H}}}}
\newcommand{\gp}{{g_{\mathrm{P}}}}
\newcommand{\omh}{{\omega_{\mathrm{H}}}}
\newcommand{\omp}{{\omega_{\mathrm{P}}}}
\newcommand{\kbT}{{k_{\mathrm{B}}T}}

\begin{document}
\title{A unification of the Holstein polaron and dynamic disorder pictures of charge transport in organic semiconductors}

\author{Jonathan H. Fetherolf}
\affiliation{Department of Chemistry and James Franck Institute, 
University of Chicago, Chicago, Illinois 60637, USA}
\author{Denis Gole\v z}
\affiliation{Center for Computational Quantum Physics, Flatiron Institute, New York, New York 10010, USA}
\author{Timothy C. Berkelbach}
\affiliation{Center for Computational Quantum Physics, Flatiron Institute, New York, New York 10010, USA}
\affiliation{Department of Chemistry, 
Columbia University, New York, New York 10027, USA}

\begin{abstract}
We present a unified and nonperturbative method for calculating spectral and
transport properties of Hamiltonians with simultaneous Holstein (diagonal) and
Peierls (off-diagonal) electron-phonon coupling. Our approach is motivated by
the separation of energy scales in semiconducting organic molecular crystals, in which
electrons couple to high-frequency intramolecular Holstein modes and to
low-frequency intermolecular Peierls modes.  We treat Peierls modes as
quasi-classical dynamic disorder, while Holstein modes are included with a
Lang-Firsov polaron transformation and no narrow-band approximation.  Our method
reduces to the popular polaron picture due to Holstein coupling and the dynamic
disorder picture due to Peierls coupling.  We derive an expression for efficient
numerical evaluation of the frequency-resolved optical conductivity based on the
Kubo formula and obtain the DC mobility from its zero-frequency component.  We
also use our method to calculate the electron-addition Green's function
corresponding to the inverse photoemission spectrum.  For realistic parameters,
temperature-dependent DC mobility is largely determined by the Peierls-induced
dynamic disorder with minor quantitative corrections due to polaronic
band-narrowing, and an activated regime is not observed at relevant
temperatures.  In contrast, for frequency-resolved observables, a quantum
mechanical treatment of the Holstein coupling is qualitatively important for
capturing the phonon replica satellite structure.
\end{abstract}

\maketitle

\section{Introduction}
Organic semiconductors are a promising class of soft materials with applications
in photovoltaics, display technologies and plastic electronics
\cite{Forrest2004, Muccini2006,Sirringhaus2014}.  In particular, ultrapure
organic molecular crystals (OMCs) provide a rich test-bed for the fundamental
mechanisms and limitations of charge transport in soft materials
\cite{Gershenson2006,Coropceanu2007}.  Exceptional room-temperature mobilities
in the tens of cm$^2$V$^{-1}$s$^{-1}$ have been measured in acene single
crystals such as rubrene and
pentacene\cite{Podzorov2004,Podzorov2005,Jurchescu2004}.  These large mobilities
are generally accompanied by an apparently ``band-like'' power-law temperature
dependence, $\mu \propto T^{-\gamma}$ ($\gamma>0$), suggesting significant
carrier delocalization over many unit cells; however controversy still exists
over the microscopic mechanisms leading to this behavior, with both experimental
and theoretical evidence suggesting an interplay between coherent dynamics,
localization, and incoherent hopping\cite{Sirringhaus2012,Karl2003}.

Theoretical efforts to understand electronic dynamics in OMCs has led to the
development of several competing and complementary pictures that capture
different limiting behaviors. For example, in the limit of weak electron-phonon
coupling, Boltzmann transport theory is quite
successful~\cite{Mahan2000,Kenkre2002,Fratini2009,DeFilippis2014,Lee2017},
similar to its use in simple inorganic semiconductors.  However, many of the
most interesting materials exhibit strong electron-phonon coupling, producing
localized electronic states and invalidating the use of Boltzmann transport
theory\cite{Fratini2009}.  In actuality, many of the material parameters that
govern charge transport in most OMCs -- the electronic bandwidth, phonon energy,
and electron-phonon coupling strength -- exist on energy scales that are
comparable to one another and to $\kbT$, precluding any simple perturbative
treatment. 

In this state of affairs, two theoretical pictures have emerged, which focus on
two distinct forms of electron-phonon coupling.  The first picture is that of
the Holstein polaron formed by strong local coupling to intramolecular
vibrations ~\cite{Holstein2000}.  In this picture, electronic motion is
generally thought of as incoherent, and nuclear tunnelling quantum effects are
non-negligible \cite{Geng2012,Nan2009a,Wang2010}.  The second picture is that of
dynamic disorder or transient localization due to strong nonlocal coupling to
intermolecular vibrations~\cite{Troisi2006a,Wang2011a,Ciuchi2011,Fratini2015a}.
In this case, the delocalized coherent evolution of the electronic wavefunction
cannot be neglected, but nuclear quantum effects are considered insignificant.
In many real OMCs, both forms of electron-phonon coupling are present.  However,
the different physical ingredients underlying these two theories prevents their
straightforward unification.  In this work, we achieve this goal and present a
unified nonperturbative approach to calculating dynamical observables in OMCs.
Importantly, our method reduces to the two powerful theories described above in
their respective limits of validity, and gives insight into the relative
importance of each energy scale and the crossover from one regime to another.
Furthermore, the method is generalizable and computationally affordable,
suggesting its straightforward future application to real materials beyond model
Hamiltonians.

The layout of this manuscript is as follows.  In Sec.~\ref{sec:theory}, we
describe the Hamiltonian and theoretical and computational tools used to
simulate dynamical properties of interest, in particular the conductivity and
the spectral function.  We further discuss the relation to previous works and
the validity of our approximations, referring to an
\hyperref[sec:appendix]{Appendix} for a comparison to numerically exact results
that we generate.  In Sec.~\ref{sec:results}, we present the results of our
simulations as a function of temperature, analyzing spectral properties and the
temperature dependence of the mobility.  Our results are presented at various
values of the strength of both types of electron-phonon coupling.  Enabled by
our unified theory, we identify a mechanism by which an increase in the
electron-phonon coupling strength counterintuitively yields a more delocalized
wavefunction.  In Sec.~\ref{sec:conc}, we summarize our results and discuss
future directions.

\section{Theory}
\label{sec:theory}

\subsection{Hamiltonian}

We study the spinless Holstein-Peierls model 
with the Hamiltonian $H = H_{\mathrm{el}} + H_{\mathrm{ph}} + H_{\mathrm{el-ph}}$
with
\begin{subequations}
\begin{align}
\label{eq:Hel}
H_{\mathrm{el}} &= \sum_{ij} h_{ij}a^{\dagger}_i a_j, \\
\label{eq:Hph}
H_{\mathrm{ph}} &= \sum_{i} \left(\omega_{\mathrm{H}}b^\dagger_{i,\mathrm{H}} b_{i,\mathrm{H}}
    + \omega_{\mathrm{P}}b^\dagger_{i,\mathrm{P}} b_{i,\mathrm{P}}\right), \\
\label{eq:Helph}
\begin{split}
H_{\mathrm{el-ph}} &= \sum_{i} g_{\mathrm{H}}\omega_{\mathrm{H}} a^{\dagger}_i a_i X_{i,\mathrm{H}} \\
    &\hspace{1em} + \sum_{\langle ij \rangle} g_\mathrm{P} \omega_{\mathrm{P}} 
        (a^\dagger_i a_j + a_j^\dagger a_i)
        (X_{i,\mathrm{P}} - X_{j,\mathrm{P}}),
\end{split}
\end{align}
\end{subequations}
where $a_i^\dagger$ ($a_i$) and $b_i^\dagger$ ($b_i$) are the creation
(annihilation) operators on lattice site $i$ for electrons
and phonons, and $\langle ij\rangle$ indicates nearest neighbors.  
Here and throughout, we use $\hbar = 1$ for notational simplicity, but all
results will be given with proper physical units.
In Eq.~(\ref{eq:Hel}), $h_{ij}$ are one-electron Hamiltonian matrix elements.
Due to particle-hole symmetry our analysis applies equally to electron and hole doping,
and in this 
work we use parameters inspired by \textit{hole} transport in rubrene\cite{Troisi2007,Girlando2010a,Girlando2011a,Ordejon2017}.
Although our ensuing formalism
can be applied to an arbitrary electronic Hamiltonian, here we will study a
nearest-neighbor tight-binding model with $h_{ij} = -\tau$ for $i$ and $j$ nearest neighbors
and zero otherwise.
This model represents one molecular orbital per lattice site and we consider two
vibrational modes on each site with frequencies $\omega_\mathrm{H}$ and $\omega_\mathrm{P}$.  
The first is a high-frequency intramolecular vibration
whose dimensionless displacement $X_{i,\mathrm{H}} \equiv (b^\dagger_{i,\mathrm{H}}+b_{i,\mathrm{H}})$ 
couples to the on-site electron density via the Holstein mechanism with dimensionless strength
$g_\mathrm{H}$. The second is a low-frequency lattice mode corresponding to the position of
the molecule in the unit cell; the difference in displacements of neighboring molecules,
$(X_{i,\mathrm{P}} - X_{j,\mathrm{P}})$, couples to the kinetic energy via the
the Peierls (or Su-Schrieffer-Heeger\cite{Su1979}) mechanism with dimensionless strength $g_\mathrm{P}$.
These two electron-phonon coupling mechanisms are commonly referred to as being local and nonlocal,
respectively.

The Hamiltonian parameters relevant for molecular organic crystals presents a
challenge to theoretical analysis.  For example, the parameters we will use in
this work are $\tau = 100$~meV, $\omega_\mathrm{H} = 150$~meV, and
$\omega_\mathrm{P} = 6$~meV.  These exemplify the typical behavior
$\omega_\mathrm{H}/\tau > 1$ and $\omega_{\mathrm{P}}/\tau < 1$, i.e.~the
intramolecular dynamics occur on timescale that is faster than the electronic
dynamics and the intermolecular vibrational dynamics occur on a timescale that
is slower than the electronic dynamics.  Furthermore, the dimensionless
electron-phonon coupling strengths $g_\mathrm{H}$ and $g_\mathrm{P}$ are both on
the order of 1 and the temperature ranges from 1 to 40~meV.  Collectively, these
parameters preclude an obvious perturbative treatment of the combined
electron-phonon interactions.  In the remainder of this section, we will discuss
our nonperturbative approach to simulating the dynamics of this Hamiltonian with
the knowledge that the Holstein and Peierls components describe fundamentally
different physical processes.

\subsection{Conductivity and mobility}
\label{sec:conductivity_theory}
Our principle object of interest is the temperature-dependent
AC conductivity $\sigma(\omega)$, which
is obtained from the current autocorrelation function,
\begin{subequations}
\begin{align}
\mathrm{Re} \sigma(\omega) &= \frac{1-e^{-\beta \omega}}{2L\omega} 
    \int_{-\infty}^{+\infty} dt\ e^{i\omega t} C_{JJ}(t),\\
C_{JJ}(t) &=
    \mathrm{Tr} [ J(t) J(0) e^{-\beta H}] / Z,
\end{align}
\end{subequations}
where $Z = \mathrm{Tr} e^{-\beta H}$ and $L$ is the number of lattice sites.  
Although in general, the conductivity is a tensor quantity with 
respect to the lattice vectors, for the sake of simplicity we will focus on the one-dimensional case.
The DC conductivity, $\sigma_{\mathrm{DC}}\equiv \sigma(\omega\rightarrow 0)$, 
gives access to the mobility via
$\mu = \sigma_\mathrm{DC}/ne_0$, where $n=N/L$ is the number density of carriers
and $e_0$ is the fundamental electric charge.
The current operator is
\begin{equation}
\label{eq:current_op}
J = -ia\sum_{\langle ij \rangle} (a^\dagger_i a_j - a_j^\dagger a_i)[-\tau+ g_\mathrm{P}\omega_{\mathrm{P}}(X_{i,\mathrm{P}} - X_{j,\mathrm{P}})]
    \equiv \sum_{ij} J_{ij} a_i^\dagger a_j
\end{equation}
and $a$ is the lattice constant and $J_{ij}$ is an operator in the Peierls phonon subspace. 
We work on a 1D lattice with
periodic boundary conditions in the canonical ensemble with $N=1$ electron,
which is appropriate for the low density limit achievable in most semiconductors
through doping or photoexcitation.

We treat the Peierls phonons quasi-classically, which is justified based on
their relatively low frequency.  Although approximate, this treatment is
nonperturbative in $g_{\mathrm{P}}$.  Under this approximation, the Peierls
phonon displacements are scalar variables that obey classical equations of
motion
\begin{equation}
X_{i,\mathrm{P}}(t) = X_{i,\mathrm{P}}(0) \cos(\omega_\mathrm{P} t)
    + P_{i,\mathrm{P}}(0) \sin(\omega_{\mathrm{P}} t)
    \label{eq:phonon_dynamics}
\end{equation}
with local Hamiltonian
$H_{i,\mathrm{P}} = (\omega_\mathrm{P}/4) 
    \left[ P_{i,\mathrm{P}}^2 + X_{i,\mathrm{P}}^2 \right]$.
This leads to a Holstein Hamiltonian with time-dependent electronic matrix elements,
\begin{equation}
\begin{split}
H_\mathrm{H}(t) &= \sum_{\langle ij \rangle}\left[ \tau_{ij}(t) \left( a_i^\dagger a_j + a_j^\dagger a_i\right)
    + \omega_\mathrm{H}b_{i,\mathrm{H}}^\dagger b_{i,\mathrm{H}} \right. \\
    &\hspace{3em} \left. + g_{\mathrm{H}}\omega_{\mathrm{H}}
        a_i^\dagger a_i X_{i,\mathrm{H}} \right],
\end{split}
\end{equation}
and a time-dependent current operator
\begin{equation}
J(t) = -ia\sum_{\langle ij \rangle} \tau_{ij}(t) (a_i^\dagger a_j - a_j^\dagger a_i)
    \equiv \sum_{ij} J_{ij}(t) a_i^\dagger a_j
\end{equation}
where
\begin{equation}
\tau_{ij}(t) = -\tau + g_\mathrm{P}\omega_{\mathrm{P}}\left[X_i(t) - X_j(t)\right].
\label{eq:tau_t}
\end{equation}
The current autocorrelation function is now approximately given by
\begin{equation}
\begin{split}
C_{JJ}(t) &= \int d\mathbf{X}_\mathrm{P} \int d\mathbf{P}_\mathrm{P}\ 
    \mathcal{P}(\mathbf{X}_\mathrm{P},\mathbf{P}_\mathrm{P}) \\
&\hspace{4em} \times
    \mathrm{Tr} \left[ U_{\mathrm{H}}(0,t) J(t) U_{\mathrm{H}}(t,0) J 
        e^{-\beta H_{\mathrm{H}}(0)} \right]/Z_{\mathrm{H}}
\end{split}
\label{eq:phase_space}
\end{equation}
where $\mathcal{P}(\mathbf{X}_\mathrm{P},\mathbf{P}_\mathrm{P})$ is the phase-space distribution of the classical 
Peierls variables and $U_\mathrm{H}(t,0)$ is the time-ordered evolution operator,
\begin{equation}
\label{eq:dd_evol}
U_\mathrm{H}(t,0) = T \exp\left[-i \int_0^t dt^\prime 
        H_\mathrm{H}(t^\prime) \right].
\end{equation}
In order to approximately respect quantum statistics at low temperature,
especially zero-point motion, the phase-space distribution $\mathcal{P}$ is a product 
of Gaussian Wigner distributions with 
$\langle X_{i,\mathrm{P}}^2 \rangle = \tanh(\beta\omega_\mathrm{P}/2)$ and 
$\langle P_{i,\mathrm{P}}^2 \rangle = 1/[\tanh(\beta\omega_\mathrm{P}/2)]$.
The integral is then evaluated by Monte Carlo sampling of $\mathbf{X}_\mathrm{P}$ 
and $\mathbf{P}_\mathrm{P}$, and these variables are then evolve in time according to classical dynamics.

To treat the Holstein electron-phonon coupling, we 
use the Lang-Firsov polaron transformation \cite{Lang1963}, defined for operators $O$ as 
$\tilde{O} \equiv e^{S} O e^{-S}$
with
\begin{equation}
S = g_\mathrm{H} \sum_i a_i^\dagger a_i (b_{i,\mathrm{H}}^\dagger - b_{i,\mathrm{H}}).
\end{equation}
Applying the transformation to $H_\mathrm{H}(t)$ and $J(t)$ gives
\begin{align}
\begin{split}
\tilde{H}_\mathrm{H}(t) &= 
    \sum_{ij} 
    \tau_{ij}(t) e^{C_i}e^{-C_j} a_i^\dagger a_j \\
    &\hspace{1em} + \sum_i \left[
    \Sigma_\mathrm{H} a_i^\dagger a_i
    + \omega_{\mathrm{H}} b^\dagger_{i,\mathrm{H}}b_{i,\mathrm{H}} \right] 
\end{split} \\
\tilde{J}(t) &= \sum_{ij} J_{ij}(t) a^\dagger_i a_j e^{C_i}e^{-C_j}
\end{align}
where $C_i = g_\mathrm{H}\left(b^\dagger_{i,\mathrm{H}} - b_{i,\mathrm{H}}\right)$
and $\Sigma_\mathrm{H} = -g_\mathrm{H}^2\omega_\mathrm{H}$ is the polaron self-energy.
We now realize a separable, finite-temperature mean-field
Hamiltonian by replacing the phonon operator $e^{C_i}e^{-C_j}$ by its thermal
average,
\begin{equation}
\begin{split}
e^{C_i}e^{-C_j} &\approx \mathrm{Tr}_\mathrm{ph} 
    \left[ e^{C_i}e^{-C_j} e^{-\beta H_{\mathrm{ph}}} \right]/Z_\mathrm{ph} \\
    &= \exp\left[-g_\mathrm{H}^2\coth(\beta\omega/2)(1-\delta_{ij})\right].
\end{split}
\end{equation}
This approximation assumes that the Holstein phonons equilibrate much faster
than the electrons, which is reasonable because of their 
high frequency $\omega_\mathrm{H} > \tau$.
We do not make any approximation in the polaronic treatment of $\tilde{J}$. 

Combining the above procedures, the total Hamiltonian has now been approximated
as a sum of separable and quadratic electron and phonon terms, where the
electronic Hamiltonian $\tilde{h}(t)$ is time- and temperature-dependent,  
\begin{equation}
\tilde{h}(t) = \sum_{\langle ij\rangle} \tilde{\tau}_{ij}(t) a^\dagger_i a_j,
\end{equation}
with the evolution operator
\begin{equation}
\tilde{u}(t,0) = T \exp\left[-i\int_0^t dt^\prime \tilde{h}(t^\prime)\right].
\end{equation}
The transfer integral is given by
\begin{equation}
\tilde{\tau}_{ij}(t) = 
    \left\{-\tau
    + g_\mathrm{P}\omega_{\mathrm{P}}\left[X_i(t) - X_j(t)\right]\right\}
    \exp\left[-g_\mathrm{H}^2\coth(\beta\omega_\mathrm{H}/2)\right]
\end{equation}
and reflects Peierls-induced dynamic disorder and Holstein-induced band narrowing.
Finally, the correlation function becomes
\begin{equation}
\label{eq:Cjjt}
C_{JJ}(t) = 
    \sum_{ijkl} J_{ijkl}(t) F_{ijkl}(t)
\end{equation}
where $J_{ijkl}(t)$ and $F_{ijkl}(t)$ are purely electronic and phononic correlation functions,
\begin{align}
\label{eq:elec_corr}
\begin{split}
J_{ijkl}(t) &= 
\int d\mathbf{X}_\mathrm{P} \int d\mathbf{P}_\mathrm{P}\ 
    \mathcal{P}(\mathbf{X}_\mathrm{P},\mathbf{P}_\mathrm{P})
J_{ij}(t) J_{kl}(0) \\
    &\hspace{2em} \times \mathrm{Tr}_{\mathrm{el}}
        \left[ \tilde{u}(0,t) a_i^\dagger a_j \tilde{u}(t,0) a_k^\dagger a_l
            e^{-\beta \tilde{h}(0)} \right] / Z_\mathrm{el}
\end{split}
\intertext{and}
\begin{split}
F_{ijkl}(t) &= \mathrm{Tr}_{\mathrm{ph}} 
    \left[ e^{C_i(t)}e^{-C_j(t)}e^{C_k}e^{-C_l} e^{\beta H_{\mathrm{ph}}} \right]/Z_\mathrm{ph} \\
    &=  \exp\left[-\tfrac{1}{2}\Phi_{ijij}(0)\right]
        \exp\left[-\tfrac{1}{2}\Phi_{klkl}(0)\right] \\
    &\hspace{1em} \times \exp\left[-\Phi_{ijkl}(t)\right],
\end{split}
\end{align}
where
\begin{subequations}
\begin{align}
\Phi_{ijkl}(t) &= \Phi(t)[\delta_{ik}-\delta_{jk}-\delta_{il}+\delta_{jl}], \\
\Phi(t) &= g_\mathrm{H}^2 [\coth(\beta\omh/2)\cos(\omh t) - i\sin(\omh t)].
\end{align}
\end{subequations}
The time dependence of $F_{ijkl}(t)$ is responsible for incoherent
electron-phonon scattering and suggests a separation into static coherent and
dynamical incoherent contributions \cite{Ortmann2009,Ortmann2010b}, $F(t) = F^\mathrm{coh} + F^\mathrm{inc}(t)$, 
with
\begin{subequations}
\begin{align}
F_{ijkl}^{\mathrm{coh}} &= \exp\left[-\tfrac{1}{2}\Phi_{ijij}(0)
    -\tfrac{1}{2}\Phi_{klkl}(0)\right] \\
F_{ijkl}^{\mathrm{inc}}(t) &= F_{ijkl}^{\mathrm{coh}} 
    \left\{ \exp\left[-\Phi_{ijkl}(t)\right] - 1\right\}.
    \label{eq:F_inc}
\end{align}
\end{subequations}
Retaining only $F^\mathrm{coh}$ produces a theory of dynamically
disordered transport with Holstein-induced band narrowing of the matrix elements
in both the Hamiltonian and the current operator.

We have implemented an algorithm for Eq.~(\ref{eq:elec_corr}) based on the time
evolution of single-particle eigenstates which, for a general electronic
Hamiltonian, scales as $L^5 N_t$ per trajectory where $L$ is the number of
lattice sites and $N_t$ is the number of timesteps.  If the electronic
Hamiltonian has only finite-range interactions, such as the nearest-neighbor
interaction used here, then the scaling is reduced to $L^3 N_t$.  Finally, if the incoherent 
phonon term is neglected, the nearest-neighbor case has a scaling of $L^2 N_t$, which is the typical
cost of the nearest-neighbor semiclassical dynamic disorder approach at finite temperature.

\subsection{One-particle spectral function}
\label{ssec:one-particle}

The above approach can be straightforwardly extended to many other correlation
functions.  In particular, we will also present results for the inverse
photoemission spectrum of the undoped parent semiconductor corresponding
to a measurement of the conduction band's many-body density of states.\cite{Mahan2000}  
This is calculated via the electron
addition Green's function of the model in the ground state with $N=0$ electrons,
\begin{equation}
iG^\mathrm{R}_{ij}(t) = \mathrm{Tr}_{(N=0)} \left[ a_i(t) a_j^\dagger e^{-\beta H} \right]/Z,
\end{equation}
using the same approximations introduced above.
From this, we calculate the momentum-resolved spectral function
\begin{equation}
A(k,\omega) = -\frac{1}{L\pi}\sum_{ij} e^{ika(i-j)}
    \mathrm{Im} \int_0^\infty dt e^{i\omega t} G^\mathrm{R}_{ij}(t)
\end{equation}
and the density of states $\rho(\omega) = L^{-1} \sum_k A(k,\omega)$.

\subsection{Limiting behaviors and relation to previous works}

In the limit $g_\mathrm{P}=0$, i.e.~without Peierls electron-phonon coupling,
our approach reduces to that presented by Ortmann et al.~\cite{Ortmann2009}  In
this limit, the Hamiltonian has no time dependence and the dynamics can be
re-written exactly in terms of eigenstates of $\tilde{h}$.  At low temperature,
this approach reduces to Boltzmann transport theory with a constant scattering
time $t_\mathrm{s}$ and band-narrowed transfer integral $\tilde{\tau}$
\cite{Ziman}.  The theory also reproduces the narrow-band and Marcus-Levich
mobility in the limit of large polaron binding energy (large $\gh$ or $\omh$),
and ultimately Marcus theory in the limit of large polaron binding energy and
high $T$ \cite{Lin2002}.  A major distinguishing feature of our approach, in
comparison to previous studies that have combined Holstein polaron approach with
dynamic disorder \cite{Wang2010}, is that we make no narrow-band approximation,
which means that we recover the exact unitary electronic dynamics in the absence
of Holstein coupling.

In the limit that $g_\mathrm{H}=0$, i.e.~without Holstein electron-phonon
coupling, this approach is analogous to the ``dynamic disorder'' picture of
Troisi and co-workers~\cite{Troisi2006a} or the transient localization scenario
of Ciuchi, Fratini and Mayou\cite{Ciuchi2011,Fratini2015a}.  In previous
work~\cite{Troisi2006a,Troisi2007,Troisi2011b,Wang2011a}, the evolution operator
Eq.~(\ref{eq:dd_evol}) is used to propagate a localized electronic state and the
mean squared displacement (MSD) is calculated; in the long-time limit, the slope
of the MSD is used to determine the diffusion coefficient and the electron
mobility.  This approach commonly includes feedback of the electronic system on
the classical phonons at a mean-field level (Ehrenfest dynamics).  We neglect
this force, resulting in the analytical solution for the phonon dynamics in
Eq.~(\ref{eq:phonon_dynamics}); this has been shown to be a safe approximation
in the medium- to high-mobility regime studied here \cite{Wang2011a}.  Ignoring
this back-reaction, our Kubo approach has three advantages.  First,
quasiclassical dynamics deteriorates in the long-time limit, approaching a
behavior that is similar to that at infinite temperature
\cite{Parandekar2005,Fratini2015a}.  Compared to the MSD, the current
autocorrelation function is less sensitive to the long-time dynamics because it
decays to zero.  Second, our approach naturally gives access to the full
frequency-resolved conductivity rather than just the DC conductivity (the
mobility).  Third, the incorporation of a fully quantum Holstein electron-phonon
interaction and its treatment by a similarity transformation is natural for the
correlation function but much harder for the wavefunction dynamics.

\subsection{Discussion of approximations}
The accuracy of our approach ultimately relies on a separation of timescales between the 
two types of phonons and the electronic system.  A semiclassical treatment of 
low-frequency Peierls modes in molecular crystals has been shown to be well-justified and
highly accurate \cite{DeFilippis2015,Fratini2015a}.  The ratio of energy scales, 
$\omp/\tau=0.06$, indicates a strongly adiabatic regime, while the temperature
$\kbT\gg\omp$ in typical experimental conditions further justifies the use of classical 
phonons.  

With respect to the Holstein modes, while the criterion of $\omh/\tau>1$ is
satisfied, this ratio (1.5 in our case) indicates that the system is only weakly
anti-adiabatic, which would limit the accuracy of our approach; many OMCs will
have a smaller bandwidth than considered here, and thus $\omh/\tau\gg1$ may be
more strongly satisfied.  To the best of our knowledge, no systematic and
detailed study has been performed that compares the Lang-Firsov approach to
numerically exact results for the Holstein model in this regime.
Therefore, in the \hyperref[sec:appendix]{Appendix}, we give a detailed 
analysis of the accuracy of the Lang-Firsov treatment compared to numerically exact calculations obtained via exact
diagonalization over the variational Hilbert space (EDVHS)~\cite{Bonca1999,Fehske2006}.  
The comparison is performed at
zero temperature, which is useful to assess accuracy even at finite temperatures, since
$\omh\gg\kbT$ for the temperature range of interest.  We find that despite
application to a challenging parameter regime, the approximate Lang-Firsov
approach captures the essential physics of the Holstein polaron in the form of
coherent band-narrowing and the incoherent electron-phonon interaction.
The accuracy of the Lang-Firsov approach holds in the limits of both weak and strong electron-phonon
coupling, demonstrating its nonperturbative nature.
Crucially, the physics enabled by our approach go well beyond those with perturbation theory and the
common narrow-band approximation.

The idea of combining classical slow phonons with quantum high-frequency modes
has been successful in the past in studies of nonadiabatic dynamics
\cite{Wang2001,Berkelbach2012b,Berkelbach2012c,Montoya-Castillo2015,Fetherolf2017,Schile2019,Teh2019}
and to a limited extent in OMC transport studies; however, such an approach
generally involves additional approximations such as static low-frequency phonons
\cite{Perroni2014} or a zero-bandwidth perturbative treatment of
high-frequency phonons\cite{Wang2010}.

\section{Results}
\label{sec:results}

\begin{figure}[t]
    \centering
    \hspace{-1mm}
    \includegraphics{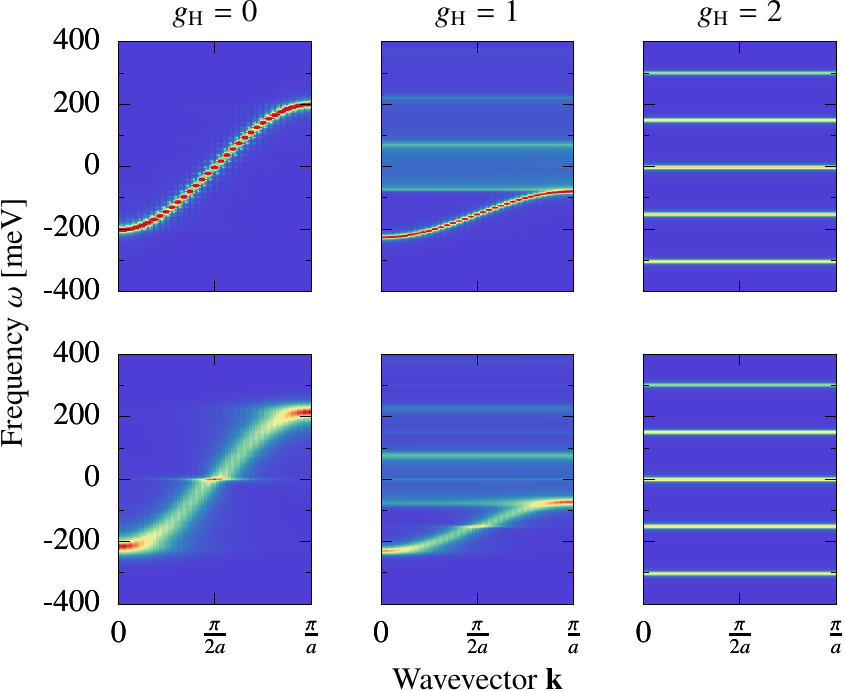}
    \caption{Momentum-resolved spectral function $A(k,\omega)$ at $T=300$~K with
no dynamic disorder (top row) and strong dynamic disorder ($\gp=2$, bottom row)
at different values of Holstein coupling $\gh$.  Calculated with electronic
transfer integral $\tau=100$~meV, phonon frequencies $\omp=6$~meV and
$\omh=150$~meV, and a Lorentzian broadening with $\eta=2.5$~meV.}
    \label{fig:Akw}
\end{figure}

\subsection{Simulation Details}
Numerical simulations were performed for a 1D periodic lattice with $L=64$ sites
for the full theory and $L=128$ sites for the pure Holstein polaron and Peierls
theories.  Phonons frequencies are $\omp=6$ meV and $\omh=150$ meV and the
electronic nearest-neighbor transfer integral is $\tau=100$ meV. These
parameters were chosen to simulate the high-mobility axis of an anisotropic
organic single crystal such as rubrene or pentacene, particularly when using a
coupling strength of $\gh=1$ and $\gp=1-2$
\cite{Troisi2007,Girlando2010a,Girlando2011a,Ordejon2017}.  In keeping with
this, we also use the lattice constant for the $b$-axis of rubrene, 7.2 $\AA$.
In addition to being representative of OMCs, these parameters satisfy the
conditions necessary to ensure the accuracy of the method presented in this
work, $\omh>\tau>\omp$, as discussed above.   Numerical propagation was
performed using the fourth-order Runge-Kutta algorithm with a timestep set to be
a factor of ten smaller than the shortest timescale in the problem,
$dt=0.1^*\min(\hbar/(4\tilde{\tau}),~1/\omh)$, which in practice varies between
0.05 and 0.4 fs.  For DC transport properties, to ensure convergence at zero
frequency in a finite system, we apply a Gaussian damping to the correlation
functions, simulating a weak dispersionless Drude scatterer such as neutral
impurity \cite{Ziman}.  We use a scattering time of $\hbar/t_{\mathrm{s}}=0.25$
meV, which is smaller than any energy scale of the problem by more than an order
of magnitude. The choice of $t_\mathrm{s}$ affects the absolute magnitude of the
mobility in cases where the current autocorrelation does not decay to zero
naturally due to other coupling, but our choice has no effect on the overall
temperature dependence.  While the spectrally-resolved quantities are obtained
from the same calculation, we instead apply an exponential damping in time
corresponding to a Lorentzian broadening in energy with half-width at half
maximum $\eta=2.5$ meV to produce smooth spectra.  Convergence with respect to
the ensemble sampling of the classical Peierls modes required averaging of up to
$10,000$ trajectories.

\subsection{Spectral functions}
Before exploring transport, it is informative to look at the one-particle
spectral properties of the Holstein-Peierls Hamiltonian within our approach,
as described in Sec.~\ref{ssec:one-particle}.  In Fig.~\ref{fig:Akw}, we
show the momentum-resolved spectral function $A(k,\omega)$ at 300~K,
with and without Peierls dynamic disorder and with various Holstein
coupling values.
In the absence of any electron-phonon coupling (top left), the spectral
function has sharp quasiparticle peaks following the expected noninteracting dispersion relation 
$E(k)=-2\tau\cos(ka)$.
When moderate Holstein coupling is added (top center), we
see a mixture of coherent and incoherent features in the spectral function.  The
zero-phonon transition is shifted by the self-energy $\Sigma_{\mathrm{H}}=-g_\mathrm{H}^2\omh$ 
and has a dispersion whose bandwidth is narrowed to $4\tilde{\tau}$.
Furthermore, satellite peaks now appear with a spacing of $\omh$, as seen in
previous results~\cite{Ranninger1993,Robin1997,Loos2006}.
With strong Holstein coupling (top right), the renormalized electronic
bandwidth is very small, leading to a dispersionless
vibronic progression with a spacing of $\omh$ and intensities 
determined by $\gh$ and $T$. 

In the case with Peierls coupling only (lower left), we see a
broadened version of the purely electronic band structure, but with an anomalous
peak at the band center.  
This unphysical peak arises in models with off-diagonal
disorder~\cite{Theodorou1976,Soukoulis1981}.
Although the peaks acquire a linewidth, the quasiparticle picture is maintained.
Dynamic disorder also yields states outside of the undisordered band
and the onset of the band edge is softened.
We recall that in the static (Anderson localization) limit, all states are
localized in one dimension; however the localization lengths of states near the band edge 
are much smaller than those in the band center \cite{Fratini2009}.  We will return to this
point in Sec.~\ref{sec:dcmob} when we analyze our mobility results in terms of localization
properties.

For the case of simultaneous Peierls and moderate Holstein coupling (bottom
center), we see that the dispersion and satellite structure largely depend on
the Holstein coupling $\gh$, whereas the Peierls dynamic disorder contributes
a broadening in frequency and momentum (as well as the spurious peak at the band
center, which is now present in the phonon replica structures).  These effects
of dynamic disorder become less pronounced as the band narrows, and in the
case with renormalized bandwidth approaching zero (bottom right), the 
Peierls disorder has no discernible effect.  

\subsection{Optical conductivity}

\begin{figure}[t]
    \centering
    \vspace{1cm}
    \includegraphics{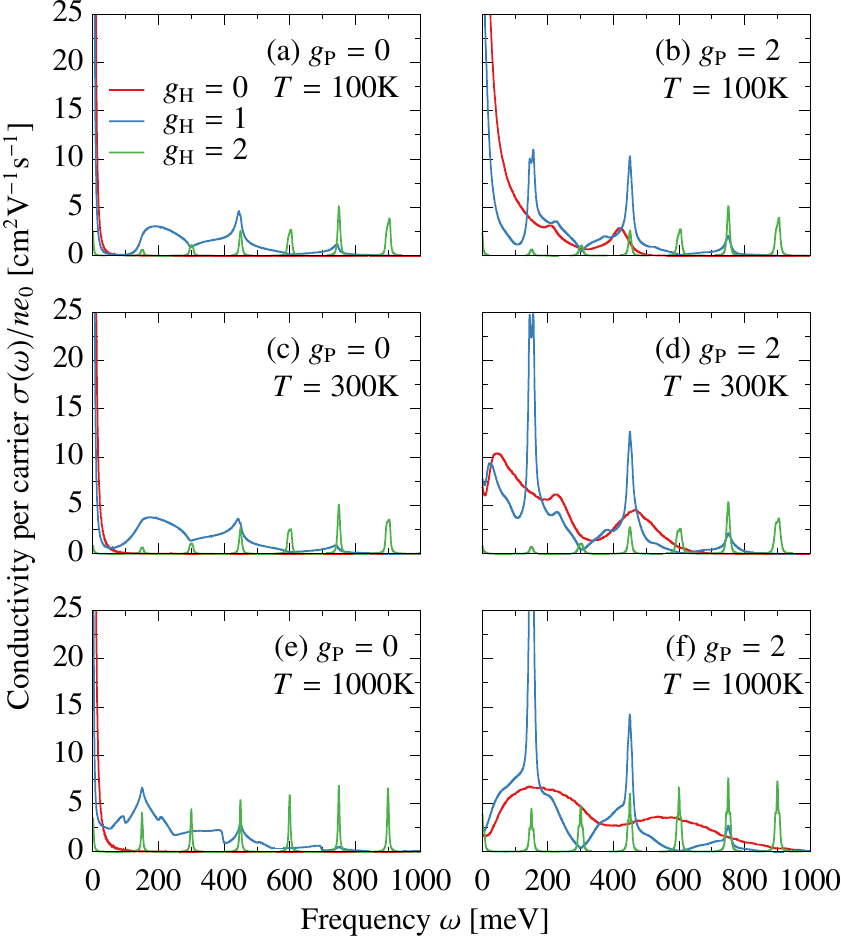}
    \caption{The per-carrier optical conductivity of the Holstein-Peierls model
at different temperatures $T=100$, 300 and 1000~K.  Results are shown for the
disorder-free case with $\gp=0$ ((a), (c) and (e)), and strong dynamic disorder
similar to that of high-mobility OMCs, $\gp=2$ ((b), (d) and (f)).  All other
parameters are the same as in Fig. \ref{fig:Akw}.}
    \label{fig:sigma_HP}
\end{figure}

We now turn our attention to the primary focus of this work, the conductivity.  In Fig. 
\ref{fig:sigma_HP} we present the AC conductivity at different temperatures for the pure (disorder-free)
Holstein model ($\gp=0$, left column) and for the full Holstein-Peierls model with strong dynamic
disorder ($\gp=2$, right column).  For the pure Holstein model, due to the aforementioned condition $\omh\gg\kbT$, we 
see little qualitative difference between the $T=100$~K and $T=300$~K cases,
Fig.~\ref{fig:sigma_HP}(a) and (c), and indeed little difference from the
zero-temperature result in Fig.~\ref{fig:sigma_exact}(a).  In order to see
qualitative changes in the pure Holstein conductivity, we need to go to much
higher temperatures ($T=1000$~K), at which point the phonon peaks associated with the
incoherent terms in Eq.~(\ref{eq:F_inc}) become more pronounced.

The full Holstein-Peierls model shows significantly different behavior.  As also seen in
the one-particle spectral function, spectra with strong Holstein coupling ($\gh=2$,
green curves) are largely unaffected by Peierls disorder.  The only qualitative
change is the appearance of weak side peaks at $\pm \omp$ at $T=1000$~K.  In all
cases with $\gh=1$ (blue curves), the addition of dynamic disorder
significantly broadens and suppresses the zero-frequency peak, while shifting
weight toward finite frequency peaks, particularly enhancing those located at
odd multiples of $\omh$.  The absence of even-numbered peaks is a coincidental
destructive interference effect, because these phonon frequencies coincide with
odd multiples of $\tau$. 
In the presence of dynamic disorder with no Holstein coupling (red
curves in Fig.~\ref{fig:sigma_HP}(b), (d) and (f)), structure is observed at 
$\omega=2\tau$ and $4\tau$, while the $\omega=0$
peak is suppressed, in agreement with previous studies \cite{Cataudella2011}.  

Summarizing our results on the finite-temperture AC conductivity of the
Holstein-Peierls Hamiltonian, the low-frequency features, which determine
transport, are most strongly affected by Peierls-induced dynamic disorder and
are virtually unaffected by the presence of moderate Holstein coupling.  
However, the Holstein coupling significantly modifies the high-frequency
structure, which can be probed spectroscopically. 
\subsection{DC mobility}
\label{sec:dcmob}
\begin{figure}[t]
    \centering
    \includegraphics{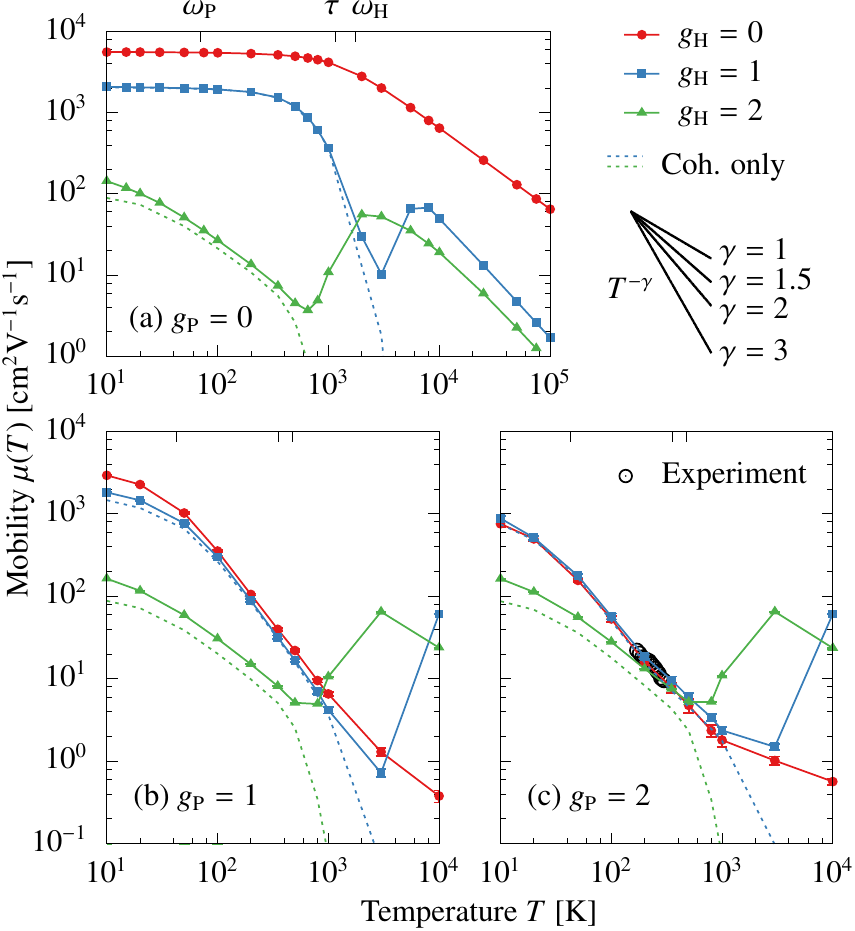}
    \caption{DC (zero-frequency) mobility with zero ($\gp=0$, (a)) Peierls
coupling, moderate Peierls coupling ($\gp=1$, (b)), and strong Peierls coupling
($\gp=2$, (c)) and various Holstein coupling.  The top axis marks the major
energies scales of the model, i.e.~Peierls frequency $\omp$, Holstein frequency
$\omh$, and the bare nearest-neighbor hopping $\tau$.  The dashed lines show the
coherent component of the mobility, where the Holstein coupling is only evident
through the renormalized electronic energy $\tilde{\tau}$.  Panel (c) compares
to rubrene $b$-axis Hall mobility measurements from
Ref.~\onlinecite{Podzorov2005}.  A Gaussian broadening of 0.25 meV was used in
all calculations to converge the conductivity at zero frequency; all other
parameters are the same as in Fig. \ref{fig:Akw}.} 
    \label{fig:DC_3panel}
\end{figure}

We now turn to a study of the DC mobility, which is the observable of greatest practical interest for 
device performance, presented in Fig.~\ref{fig:DC_3panel}.
First, we recall the single-carrier ($N=1$) mobility in the absence of any electron-phonon coupling, corresponding
to the Boltzmann transport equation~\cite{Ziman,Mahan2000,Ortmann2009}
\begin{equation}
\mu_{\mathrm{el}} = \frac{\sqrt{\pi}t_\mathrm{s}}{e_0 k_B T Z_{\mathrm{el}}} \Tr{ J^2 e^{-\beta h}}
    = \frac{\sqrt{\pi}e_0t_\mathrm{s}}{ \kbT Z_\mathrm{el}} \sum_k v^2_k e^{-\eps(k)/k_\mathrm{B}T},
\label{eq:BTE}
\end{equation}
where $t_\mathrm{s}$ is a Drude scattering time and $v_k = \nabla_k \eps(k)$.  
At low temperatures over which the band is effectively parabolic, a single electron in the canonical ensemble
obeys ideal statistics
$\langle v^2 \rangle = \kbT/m^*$, producing a constant value
\begin{equation}
    \mu_\mathrm{el}(T\ll\tau) = \frac{\sqrt{\pi} e_0 t_\mathrm{s}}{m^*}
    = 2\sqrt{\pi} e_0 a^2 t_\mathrm{s} \tau, 
    \label{eq:drude}
\end{equation}
which is the Drude mobility.  At high temperatures, $\kbT \gg \tau$, the average over the finite
bandwidth is constant, $\langle J^2 \rangle \approx 4\pi e_0^2 \tau^2 a^2$, giving
\begin{equation}
\mu_\mathrm{el}(T\gg\tau) = 4\pi^{3/2} e_0 a^2 t_\mathrm{s} \frac{\tau^2}{\kbT},
\end{equation}
i.e.~decaying like $T^{-1}$. 
The crossover between these two behaviors clearly occurs at $\kbT \approx \tau$,
as can be seen in Fig.~\ref{fig:DC_3panel}(a), with $\gh,\gp=0$ (red data).

With the addition of Holstein coupling, our theory of the mobility reduces to that
of Ortmann et al.~\cite{Ortmann2009} and our analysis is correspondingly similar.  
The coherent contribution to the mobility has the same form as above, but with
the temperature-dependent renormalized $\tilde{\tau}(T)$. 
Thus the low-temperature constant value is reduced
by a factor of $e^{-g^2_{\mathrm{H}}}$, the crossover temperature is reduced, and
the high-temperature decay is faster than $T^{-1}$.  The mobility has an
additional incoherent contribution that yields activated behavior at
intermediate temperature, where the activation temperature decreases with
increasing $\gh$.
At very high temperatures $\kbT \gg g^2\omh$, we see the
well-known $T^{-3/2}$ power-law dependence of Marcus theory \cite{Marcus1956}.
In Fig.~\ref{fig:DC_3panel}(a), we extend our results to unrealistically high
temperatures $T \sim 10^5$~K only to show this asymptotic behavior.

\begin{figure}[t!]
    \centering
%    \vspace{1cm}
    \includegraphics{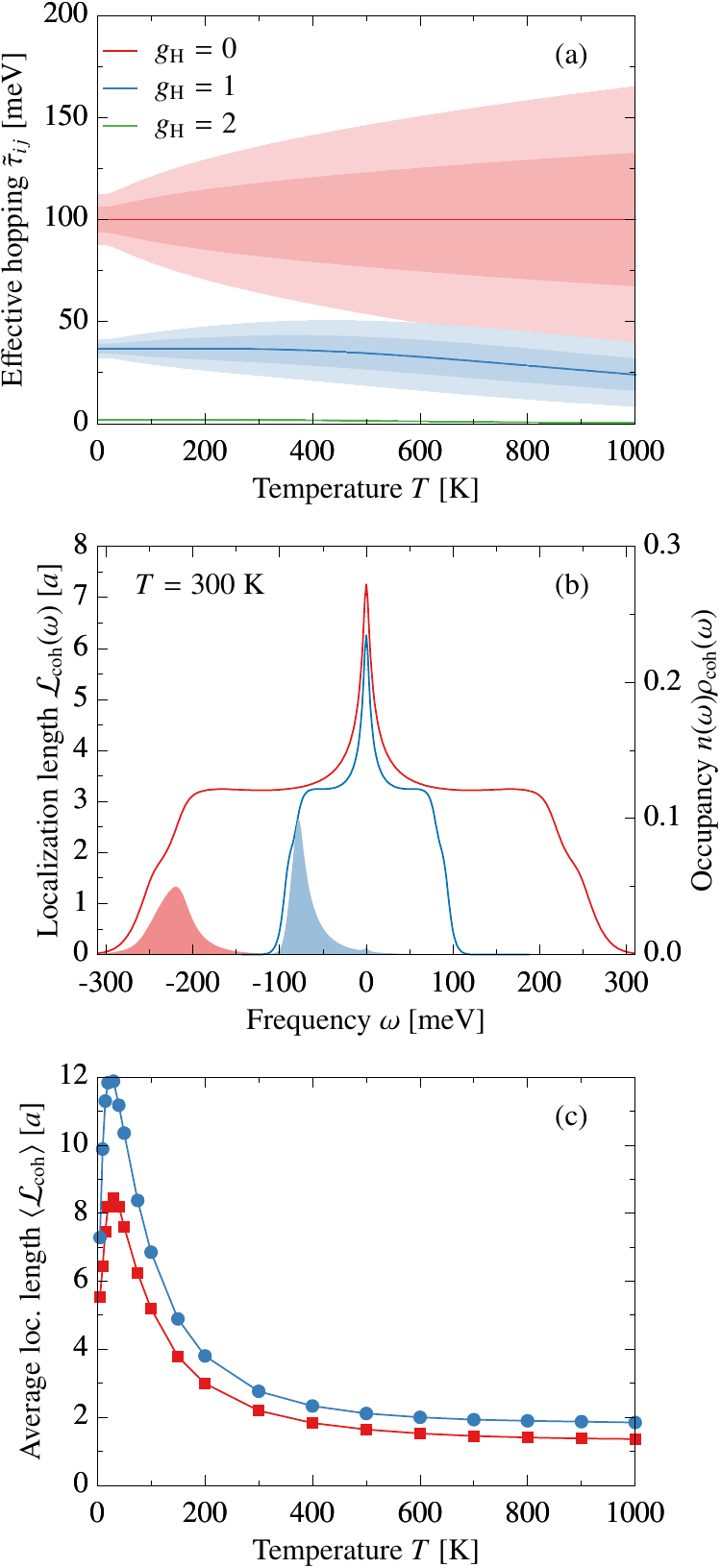}
    \caption{(a) Band-narrowed nearest-neighbor hopping $\tilde{\tau}$ (solid
lines), for which the bare value is $\tau=100$ meV.  Shaded regions represent
one standard deviation in the distribution of values that $\tilde{\tau}$ can
take under Peierls disorder with moderate coupling ($\gp=1$, darker shading) and
strong coupling ($\gp=2$, lighter shading).  
(b) Energy-dependent localization length (solid lines) for zero Holstein coupling ($\gh=0$, red)
and moderate Holstein coupling ($\gh=1$, blue) and thermally occupied density of states (shaded regions).
Results are shown for $T=300$~K and Peierls coupling $\gp=2$. 
(c) Temperature-dependent average
localization length  with Peierls coupling $\gp=2$.  All other parameters are the same as in Fig. \ref{fig:Akw}.}
    \label{fig:locT}
\end{figure}

Aside from the effects of band-narrowing -- an overall reduction of mobility and
faster than $T^{-1}$ decay -- the major signature of a high-frequency
Holstein phonon is the presence of the activated regime.
However, our numerical results indicate that for parameters relevant to a
high-mobility OMC, the activated regime is only observed at 
temperatures exceeding 1000~K.  This is due to the high frequency of the 
intramolecular vibrations, on the order of $150-200$~meV.  
Therefore, we conclude that an intrinsic activated regime associated with polaron formation
may not be observable in OMCs.  However, this does not indicate the absence 
nor irrelevance of strong intramolecular electron-phonon coupling.

Lastly, we analyze the DC mobility in the presence of both Holstein and Peierls 
electron-phonon couplings, which is only enabled by the new theory developed here.
In comparing Fig. \ref{fig:DC_3panel}(b) and (c) (moderate and strong
Peierls coupling, $\gp=1$ and $\gp=2$) to (a) (no Peierls coupling), it is clear that the
temperature dependence of the mobility is dramatically affected by dynamic
disorder, except for the case of very strong Holstein coupling (green
data).  The reason for this latter insensitivity
is illustrated in Fig. \ref{fig:locT}(a), which shows the mean value
(solid lines) and standard deviation of the renormalized transfer integral under moderate ($\gp=1$) and strong
($\gp=2$) Peierls disorder (dark and light shaded regions, respectively).  We
see that for $\gh=2$, the mean value is nearly zero and, concomitantly, 
the fluctuations with these values of $\gp$ are negligible.

Having seen that Peierls dynamic disorder has little effect in the large $\gh$ limit, we will mainly compare 
the $\gh=0$ and $\gh=1$ cases.  As previously stated, $\gh=1$ corresponds most
closely to the estimated coupling strength in rubrene \cite{Girlando2010a,Girlando2011a,Ordejon2017,Troisi2007},
whereas the $\gh=0$ case reduces to the dynamic disorder/transient
localization picture that has so successfully captured the transport behavior of
high-mobility OMCs in the past\cite{Troisi2006a,Fratini2015a}. In the temperature range of interest
($10^2$-$10^3$K), the $\gh=0$ and $\gh=1$ Holstein-Peierls mobilities
are remarkably similar, exhibiting the well-known $\mu \propto T^{-\gamma}$ behavior, with $\gamma \approx 2$
characteristic of dynamic disorder-induced transient localization \cite{Fratini2009,Ciuchi2011,Fratini2015a}.  
For $\gp = 2$ and $\gh=0$ or 1, our extracted value of $\gamma \approx 1.8$ is slightly
smaller than that determined from the slope of the mean-squared displacement in Refs.~\onlinecite{Troisi2006a,Troisi2007}.
This effect is potentially related to the spurious heating of quasiclassical
dynamics\cite{Ciuchi2011,Parandekar2005} and which is significantly less
influential in our formulation based on the Kubo formalism. 

The only major qualitative change induced by Holstein coupling is the activated
behavior, which again occurs at unrealistically high temperatures exceeding
1000~K.  In the dotted lines of Fig.~\ref{fig:DC_3panel}, we show the mobility
calculated in the absence of incoherent phonon effects, which is completely
sufficient for experimentally relevant temperatures.  This observation argues
that the mobility of many OMCs should be simulated by a computational approach
where the transfer integrals undergo a static but temperature-dependent
renormalization due to the Holstein coupling to high-frequency intramolecular
vibrations. These parameters are then used in a quasiclassical, time-dependent
simulation of the electronic dynamics, following the usual prescriptions of
dynamic disorder\cite{Troisi2006a,Troisi2007,Wang2011a} or transient
localization \cite{Fratini2009,Ciuchi2011,Fratini2015a}.  In Fig.
\ref{fig:DC_3panel}(c), we compare our theoretical results to experimental Hall
mobility data of the rubrene $b$-axis via Ref.~\onlinecite{Podzorov2005}.  Both
the $\gh=0$ and $\gh=1$ data show excellent agreement in terms of the absolute
magnitude and temperature dependence of the mobility.

Before concluding, we seek to understand the origin of the mobility's insensitivity to the Holstein coupling
strength $\gh$, in cases where the Peierls coupling is nonnegligible.
Figure~\ref{fig:locT}(a) clearly shows that the magnitude of the renormalized transfer integral is significantly
affected by $\gh$, which would suggest a strong dependence.  However, Fig.~\ref{fig:locT}(a) also demonstrates that
the band-narrowing reduces the variance of the transfer integrals, which lessens the degree of localization.
To test the proposal, we analyze the localization properties of the 
electronic wavefunction averaged over disorder realizations and 
with statically renormalized transfer integrals.
The energy-resolved coherent localization length, 
$\mathcal{L}_\mathrm{coh}(\omega)$ can be
obtained via the Thouless formalism \cite{Thouless1972,Li2013},
\begin{subequations}
\begin{align}
\mathcal{L}^{-1}_\mathrm{coh}(\omega)
    &= \frac{1}{L} \sum_\alpha \frac{l^{-1}_\alpha \delta(\omega - \tilde{\varepsilon}_\alpha)}{\rho_\mathrm{coh}(\omega)} \\
l_\alpha^{-1} &= \frac{1}{(L-1)a}
    \left(\sum_{\beta\neq\alpha}\ln|\tilde{\varepsilon}_\beta
        - \tilde{\varepsilon}_\alpha|-\sum_{i=1}^{L-1}\ln|\tilde{\tau}_{i,i+1}| \right) \\
\rho_\mathrm{coh}(\omega) &= \frac{1}{L} \sum_\alpha \delta(\omega-\tilde{\varepsilon}_\alpha)
\end{align}
\end{subequations}
where $\tilde{\varepsilon}_\alpha = \tilde{\varepsilon}_\alpha(\mathbf{X}_\mathrm{P},\mathbf{P}_\mathrm{P})$ are eigenvalues of the 
band-narrowed and statically disordered electronic Hamiltonian.
The above quantities $\mathcal{L}_\mathrm{coh}(\omega)$ and $\rho_\mathrm{coh}(\omega)$ are thermally averaged over the
phonon degrees of freedom by Monte Carlo sampling as done in Eq.~(\ref{eq:phase_space}).

In Fig.~\ref{fig:locT}(b), we show this disorder-averaged energy-dependent
localization length $\mathcal{L}_\mathrm{coh}(\omega)$ without and with
Holstein electron-phonon coupling ($\gh=0$ and $\gh=1$) at an example temperature of
$T=300$~K.  We observe three main
features: highly localized states at the band edges, states with a constant
localization length of a few lattice constants in the middle of the band, and an
unphysical delocalized state at the band center due to the purely off-diagonal
nature of the disorder.  Crucially, in the case with Holstein electron-phonon
coupling ($\gh=1$), all of these features are compressed into a narrower energy
spacing.
The shaded region of Fig.~\ref{fig:locT}(b) shows the thermally occupied density of
states $n(\omega)\rho_\mathrm{coh}(\omega)$, where $n(\omega)$ is the thermal
occupancy.  We see that the $\gh=1$ occupied density of states extends much
further into the band, giving added weight to the more delocalized states.
This yields a \textit{larger} average localization length, which is defined
by
$\langle \mathcal{L}_\mathrm{coh} \rangle = \int d\omega n(\omega) \rho_\mathrm{coh}(\omega) \mathcal{L}_\mathrm{coh}(\omega)$
and plotted in Fig.~\ref{fig:locT}(c).
Indeed, at all temperatures we see that the average localization length for a system with $\gh=1$ is \textit{larger}
than that with $\gh=0$.  In other words, the addition of Holstein electron-phonon coupling yields electronic states
that are, on average, \textit{more} delocalized.
This parameter-specific effect increases the mobility and partially compensates for the decreased magnitude of the renormalized
transfer integrals, producing a mobility that is relatively insensitive to the value of $\gh$, as seen in Fig.~\ref{fig:DC_3panel}.

More broadly, at all values of $\gh$, we observe a maximum delocalization at low
temperature and exponential decay with increasing temperature, as previously
observed by Ciuchi and Fratini \cite{Fratini2009}.  In our model, maximum
delocalization does not occur at zero temperature because we include zero-point
energy in the Peierls phonons by sampling from the Wigner distribution, which
allows for localized states even at zero temperature.  Interestingly, at these
low temperatures $\kbT < \omp$ (with Wigner sampling), the degree of disorder is
temperature-independent.  The temperature dependence of the localization length
only comes from the thermal average over electronic states, where higher energy
states have a larger localization length, causing the average localization
length to increase with increasing temperature up to $\kbT\approx \omp$.  To
summarize, we have seen that the temperature-dependent mobility is determined by
a subtle competition between a Holstein-induced renormalization of electronic
parameters and their variance, the finite-temperature dynamics of quasiclassical
Peierls modes, and the thermal occupancy of electronic states with different
localization lengths.

\section{Conclusions}
\label{sec:conc}
We have introduced a new approach to solving the Holstein-Peierls Hamiltonian
for electron-phonon coupling in organic molecular crystals.  We exploit the
quasi-adiabatic nature of the intermolecular Peierls modes and treat them
semiclassically.  The lattice motions create a dynamically disordered landscape
with broken translational symmetry that localizes the polaronic wavefunction.
The intramolecular Holstein modes are accounted for by performing a Lang-Firsov
polaron transformation, and separating the electron and Holstein phonon degrees
of freedom through an anti-adiabatic finite-temperature mean-field approximation. The resulting
dynamics are nonperturbative in both the electronic and electron-phonon
interactions.  The polaronic dynamics exactly capture the limit of strong
electron-phonon coupling, while reducing to the exact unitary electronic
dynamics in the weak-coupling limit.  We calculated the frequency-resolved
spectral function and conductivity, as well as the zero-frequency DC mobility.

We found that finite-frequency observables such as the spectral function and 
optical conductivity are strongly affected by Holstein phonons even at 
moderate coupling strengths, especially at frequencies corresponding to 
multiples of the phonon energy.  These features correspond to incoherent 
electron-phonon interactions, and are only enabled by a quantum mechanical 
treatment such as that presented here.  Peierls disorder contributes 
peak-broadening and shifts spectral weight toward incoherent peaks.  DC 
transport, in contrast to finite-frequency observables, is largely determined
by dynamic disorder-induced localization of the coherent 
wavefunction~\cite{Troisi2006a,Wang2011a,Ciuchi2011,Fratini2015a}.  The DC 
mobility is only weakly affected by the Holstein interaction in the form of
temperature-dependent band-narrowing, i.e.~effective mass renormalization.  
An incoherent activated 
regime due to Holstein phonons is observed, but only at experimentally irrelevant 
temperatures, at least for the frequency of intramolecular vibrations considered here
and relevant for most OMCs.  We 
further support this conclusion by showing that including only Holstein 
band-narrowing, and excluding incoherent features, can reproduce the total mobility to high accuracy up to temperatures of around 1000~K.  

Interestingly, 
we find that coherent band-narrowing due to Holstein coupling has less of an 
effect on overall transport than might be expected, because the bandwidth 
reduction is accompanied by a reduction in the variance and thus a reduction in
the nonlocal dynamic disorder.  Such an observation is only made possible by the
unified theory presented in this work.  This finding further
indicates that the success of theories with purely nonlocal electron-phonon coupling 
may be partially due to a serendipitous cancellation of effects which makes 
local coupling to intramolecular vibrations appear less significant.  Previous
studies have shown that dynamic disorder enhances hopping transport in 1D, but
the mechanism was due to completely incoherent pathways and therefore unrelated
to localization \cite{Wang2010}.  Because localization properties are strongly
dependent on dimensionality and potential anisotropy \cite{Fratini2017}, the interplay of
localization and band-renormalization may be significantly different in higher
dimensions, which we consider an interesting topic for future study.

Perhaps most importantly, the relative simplicity of our approach enables application to more complex
models and realistic materials, where numerically exact techniques struggle.  
Because the theory leads to a
time-dependent Hamiltonian with separable electronic and phononic degrees of freedom, 
we imagine the straightforward inclusion of electronic interactions, treated using conventional
methods of electronic structure theory in the absence of phonons.  This extension would be important
in semiconductors at higher doping densities or in metals with strong electron-phonon interactions.
Furthermore, a more sophisticated treatment of the electron-phonon
interaction, for example via variational optimization of the Lang-Firsov
transformation, is also possible \cite{Yarkony1977,Perroni2012,Prodanovic2019}. 
Finally, we emphasize that the quasiclassical treatment of low-frequency nuclear motion is 
not limited to the study of linear electron-phonon coupling or to 
harmonic nuclear degrees of freedom.  Fully atomistic and anharmonic nuclear dynamics can be
straightforwardly included; anharmonic effects are most important for low-frequency
motions, which is precisely where the approach is valid.
For example, in addition to OMCs \cite{Ruggiero2017}, nonlinear electron-phonon coupling and anharmonic effects
have been shown to be important in the electronic properties of lead-halide
perovskites \cite{Yaffe2017a,Mayers2018} and for ``phononic'' control of nonequilibrium material
properties \cite{Nicoletti2016,Mankowsky2014}.  
Work along all of these directions is underway and will enable nonperturbative treatment
of electron-nuclear interactions in realistic, complex materials.

\section*{Acknowledgements}
We thank Omer Yaffe for helpful discussions.
This research was supported by the US-Israel Binational Science Foundation Grant
BSF-2016362 and by start-up funds from the University of Chicago.  J.H.F. was
supported in part by the National Science Foundation Graduate Research
Fellowship Grant DGE-1746045.  Calculations were performed with resources
provided by the University of Chicago Research Computing Center (RCC).  The
Flatiron Institute is a division of the Simons Foundation.

\renewcommand\thesubsection{\arabic{subsection}}
\setcounter{subsection}{0}
\section*{Appendix: Validity of the approximate Lang-Firsov treatment}
\label{sec:appendix}
\setcounter{equation}{0}
\renewcommand\theequation{A\arabic{equation}}

In this appendix, we seek to benchmark the accuracy of the approximate
Lang-Firsov treatment of the Holstein model by comparison with numerically exact
results, using the same material parameters as considered in the main text.  A
large number of works have been devoted to the numerically exact treatment of
electron-phonon problems, especially the Holstein model.  In particular,
variational techniques~\cite{Shi2018}, the density matrix renormalization
group~\cite{zhang1999,Brockt2015}, exact diagonalization of small
clusters~\cite{Bonca1999,Fehske2006} and diagrammatic
techniques\cite{berciu2006} have been used to predict highly accurate or exact
spectral properties for modestly sized 1D systems.  Higher dimensional
extensions have been explored within exact diagonalization\cite{Ku2002} and
dynamical mean field theory~\cite{werner2007}. The later has been extended to
finite doping and symmetry-broken phases~\cite{murakami2013,murakami2014}.
Here, we employ exact diagonalization over the variational Hilbert
space~(EDVHS), which is described in the next section.

\subsection{Exact diagonalization over variational Hilbert space}
\label{sec:app_EHVHS}
EDVHS can be used to obtain a numerically exact solution of the Holstein polaron
ground state\cite{Bonca1999}, response functions~\cite{Fehske2006} and
nonequilibrium dynamics\cite{Vidmar2011,Golez2012}. The construction of the
variational space starts from an electron in a translationally invariant state
with no phonon excitations. We generate new parent states by applying the
Hamiltonian, Eqs.~(\ref{eq:Hel}) and (\ref{eq:Helph}). By choosing the number of
Hamiltonian operations $N_\mathrm{H}$, we control the maximal number of phonon
quanta and the spatial extent of the polaron, which can be increased for
convergence.  After solving the ground state problem, the dynamical properties
are evaluated using the continued fraction method\cite{Dagotto1994}.

All EDVHS calculations are converged with $N_{\mathrm{H}}=20$ generations
creating a Hilbert space with over $3\times10^6$ basis states.  Lorentzian
broadening with $\eta=10$~meV for spectral functions and 15 meV for optical
conductivity spectra were used in the presentation of spectra. 

\subsection{Spectral functions}
\label{sec:app_Akw}
\begin{figure}[t]
    \centering
    \hspace{-1mm}
    \includegraphics{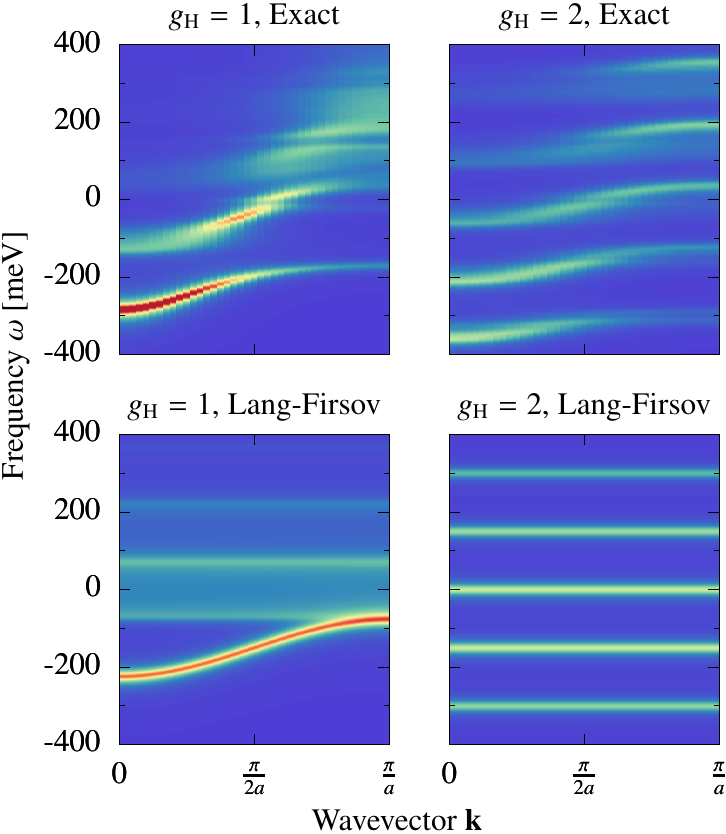}
    \caption{Momentum-resolved spectral function $A(k,\omega)$ at
    zero temperature obtained by numerically exact EDVHS results
    (top row) and approximate Lang-Firsov treatment (bottom row) at two
    different values of the Holstein electron-phonon coupling constant $\gh$.  
    Results are calculated with the
    electronic transfer integral $\tau=100$~meV, Holstein phonon
    frequency $\omh=150$~meV, and a Lorentzian broadening with
    $\eta=10$~meV.
    }
    \label{fig:Akw_exact}
\end{figure}

The momentum-resolved spectral function for the Holstein model at zero
temperature is presented in Fig.~\ref{fig:Akw_exact}.  We compare the
numerically exact EDVHS results (top row) to the approximate Lang-Firsov
treatment (bottom row).  For intermediate coupling ($\gh=1$) we see good
agreement, particularly in the dispersion of the quasiparticle peak and in the
satellite structure near $k=0$.  These features are most relevant to the
transport properties, which rely on the $k=0$ contribution of the two-particle
Green's function \cite{Mahan2000}.  The satellite peaks, which are strictly
dispersionless in the Lang-Firsov treatment, inherit some of the electronic
dispersion in the exact results.  At stronger electron-phonon coupling ($g=2$),
we see better agreement in the vibronic replica structure at all values of $k$;
however the exact result again has slight dispersion in the peaks that is absent
in the approximate result.

\begin{figure}[t]
    \centering
    \hspace{-1mm}
    \includegraphics{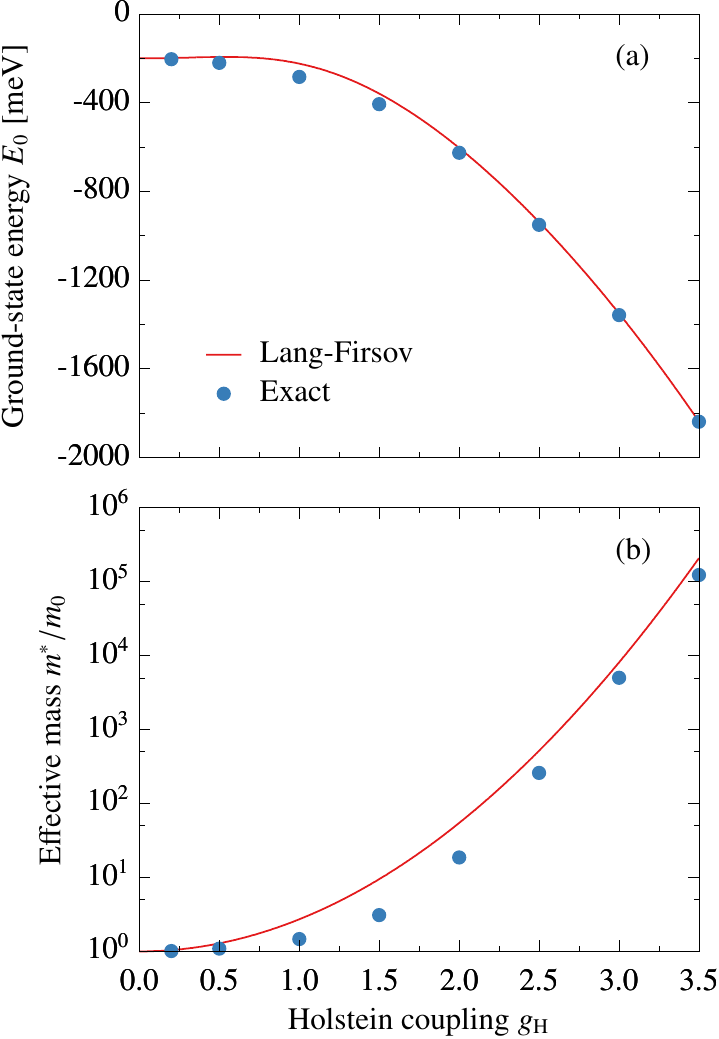}
    \caption{(a) Ground-state energy $E_0$ and (b) effective mass enhancement $m^*/m_0$ of the Holstein polaron at zero temperature, calculated via approximate Lang-Firsov and exact EDHVS.  Calculated for the weakly anti-adiabatic case where $\tau=100$ meV and $\omh=150$ meV.}
    \label{fig:meff}
\end{figure}

The positions of the spectral peaks are slightly shifted in the approximate
result, which is due to an inaccurate ground-state energy of the $N=1$ polaron
problem.  We examine this ground-state energy as a function of coupling strength
$\gh$ in Fig.~\ref{fig:meff}(a), which shows that the Lang-Firsov approach is
accurate and both weak and strong coupling; the intermediate-coupling regime of
$\gh=1-2$ is most challenging, but is qualitatively captured.
We also note that a constant energy shift is not relevant for the conductivity,
which depends only on energy differences at fixed electron number.

More important for transport is the quasiparticle effective mass at the bottom of the band.
In Fig. \ref{fig:meff}(b) we look at the effective mass enhancement
\begin{equation}
    \frac{m^*}{m_0}=\bigg[\frac{1}{2\tau}\nabla_k^2E(k)\Big|_{k=0}\bigg]^{-1},
\end{equation}
which, within the Lang-Firsov approximation, is the inverse of the
band-narrowing factor; at zero temperature, $m^*/m_0=\tau/\tilde{\tau}=
e^{\gh^2}$.  Again we see that the Lang-Firsov approach is accurate at small and
large values of the coupling constant but slightly overestimates the mass
enhancement at intermediate coupling.  

Overall, we see excellent agreement between the approximate and 
exact spectral functions even in the weakly anti-adiabatic, 
intermediate-coupling regime relevant to high-mobility organic 
semiconductors.  The approximate method is particularly accurate 
for the low-momentum, low-energy features that are relevant for 
conductivity and DC transport at low density.

\subsection{Optical conductivity}
\label{sec:app_sigma}
\begin{figure}[t]
    \centering
    \vspace{1cm}
    \includegraphics{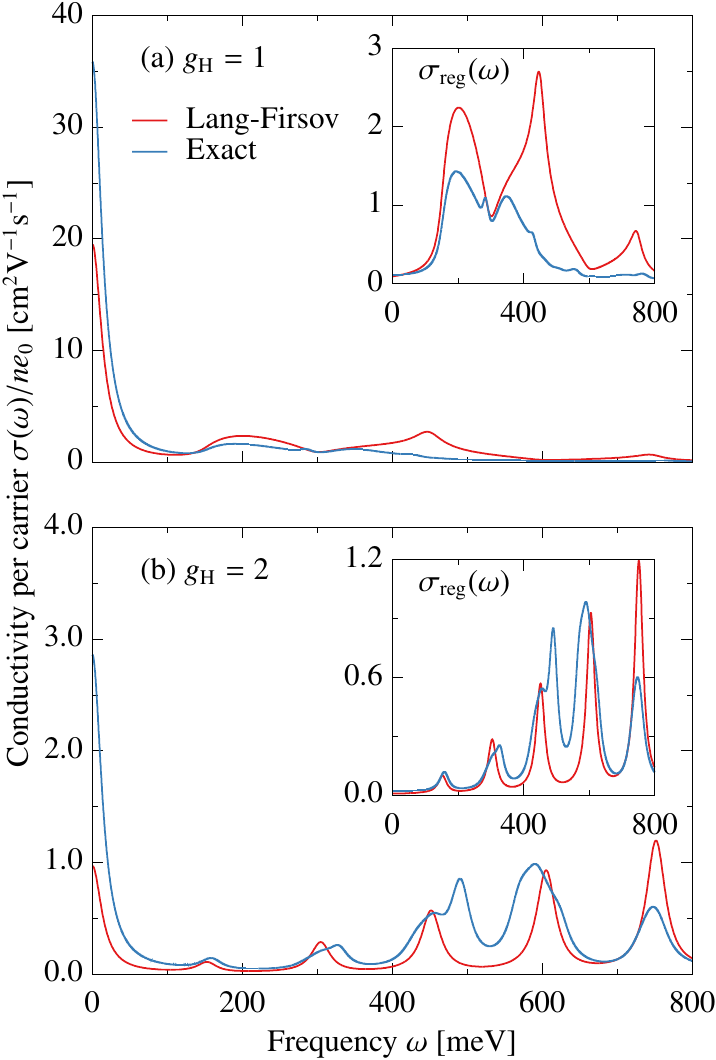}
    \caption{The per-particle optical conductivity $\sigma(\omega)/ne_0$ of the Holstein model
     at zero temperature obtained by numerically exact EDVHS and approximate Lang-Firsov treatment.
    Insets show the
    non-Drude regular conductivity, 
    $\sigma_{\mathrm{reg}}(\omega)=\sigma(\omega)-D\delta(\omega)$.  Parameters are the same
    as in Fig.~\ref{fig:Akw_exact} except the broadening is $\eta = 15$~meV. 
    Note that the vertical scales of (a) and 
    (b) differ by a factor of 10.
    }
    \label{fig:sigma_exact}
\end{figure}

Next we examine the frequency-dependent optical conductivity, $\sigma(\omega)$, which is presented in 
the main panels of Fig.~\ref{fig:sigma_exact}.  The conductivity can be separated into 
a zero-frequency Drude contribution $D\delta(\omega)$ and a regular finite-frequency
contribution $\sigma_{\mathrm{reg}}(\omega)$; the latter is shown in the insets.  
Overall, the structure of the conductivity is well-reproduced by the Lang-Firsov treatment, including peak locations and lineshapes.
The exact results show a complex mixture of 
coherent and incoherent features that contributes fine structure, which is absent in the approximate spectrum.  
The approximate Lang-Firsov approach predicts a structure that is 
dominated by regularly spaced vibronic replicas similar to that seen in the single-site limit~\cite{Mahan2000}.

The low-frequency behavior, which determines transport properties, is qualitatively reproduced by the
approximate treatment.  For the case with moderate coupling ($\gh=1)$,
the Drude weights are in satisfactory
agreement: $D=0.68$ (exact) and $D=0.37$ (approximate).
At strong coupling ($\gh=2$), the agreement is slightly worse: $D=0.054$ (exact) and $D=0.018$ (approximate).
At low density, the Drude weight is related to the effective mass via $D \propto 1/m^*$ and the underestimation of the
Drude weight is consistent with the overestimation of the mass enhancement seen in Fig.~\ref{fig:meff}(b).

To summarize our comparison of the conductivity at zero temperature, the
low-frequency features show reasonable accuracy, including the zero-frequency
Drude contribution and the locations and lineshapes of the first few
excited-state peaks.  Crucially, the nonperturbative Lang-Firsov approach
captures exactly the weak- and strong-coupling limits, enabling far richer
physics than those weak-coupling perturbation theory or the historically
prevalent narrow-band approximation.

\end{document}